\documentclass[aps,prb,twocolumn,showpacs,amsmath,amssymb,superscriptaddress]{revtex4-2}
\usepackage{graphicx}
\usepackage{dcolumn}
\usepackage{amsmath}
\usepackage{amssymb}
\usepackage{color}
\usepackage{epstopdf}
\newcommand{\erfc}{\mathrm{erfc}}
\begin{document}
\title{ Polarization angle dependence of the breathing modes in confined one-dimensional dipolar bosons}
\author{S. De Palo}
\affiliation{CNR-IOM-Democritos National Simulation Centre, UDS Via Bonomea 265, I-34136, Trieste, Italy}
\affiliation{Dipartimento di Fisica Teorica, Universit\`a Trieste, Strada Costiera 11, I-34014 Trieste, Italy}
\author{E. Orignac}
\affiliation{Univ Lyon, Ens de Lyon, Univ Claude Bernard, CNRS, Laboratoire de Physique, F-69342 Lyon, France}
\author{M.L. Chiofalo}
\affiliation{Dipartamento di Fisica, Universit\`a di Pisa, Italy}
\author{R. Citro}
\affiliation{Dipartamento di Fisica ``E. R. Caianiello'', Universit\`a
  degli Studi di Salerno, Via Giovanni Paolo II, I-84084 Fisciano (Sa), Italy}
\begin{abstract}
Probing the radial collective oscillation of a trapped quantum system is an accurate
experimental tool to investigate interactions and dimensionality effects.
We consider a fully polarized quasi-one dimensional dipolar quantum gas of bosonic dysprosium atoms in a parabolic trap
at zero temperature. We model the dipolar gas with an effective quasi-one dimensional
Hamiltonian in the single-mode approximation, and derive the equation of state using a  variational approximation based on the Lieb-Liniger gas Bethe Ansatz
wavefunction or  perturbation theory. We calculate the breathing mode frequencies while varying polarization angles by a sum-rule approach, and find  them
in good agreement with recent experimental findings.
\end{abstract}
\date{\today}
\maketitle
\section{Introduction}
\label{sec:intro}


Systems with long-range interactions present a host of exotic quantum states of matter including  Wigner crystals~~\cite{schulz_wigner_1d,capponi_coulomb_1dmeso}, Haldane Insulators~~\cite{dalla_torre2006} or Fulde-Ferrell-Larkin-Ovchinnikov phases~\cite{wei_fflo_2018}, thanks to the interplay
between quantum fluctuations and the frustrating effects of interactions. In particular, the advent of degenerate quantum gases consisting of atoms where strong dipolar forces provide the interactions, has even revealed the coexistence of both crystalline order and superfluidity, the \emph{so-called} supersolidity~~\cite{PhysRevLett.122.130405,PhysRevX.9.011051,PhysRevX.9.021012}.

Recently, the possibility of forming one-dimensional tubes of bosonic $Dy$ atoms with tunable
strength of the contact and dipolar interactions~~\cite{tang2018} has opened the fascinating perspective of
investigating the interplay between quantum fluctuations, enhanced in reduced dimensionality, and interaction-driven fluctuations, leading to unconventional relaxation mechanisms and the so-called scar states~~\cite{Lev2020}.
In fact, although in one dimension repulsive dipolar interaction decaying as $1/r^3$ at long distance are classified as finite-ranged interaction, they are expected to push bosonic systems to a regime of stronger interactions~~\cite{citro06_dipolar1d,citro07_luttinger_hydrodynamics,Roscilde2010}.


Since the majority of existing ultracold-gas experiments are carried out
with  spatially inhomogeneous systems, due to
the presence of an external confining potential,  exciting
oscillations of the gas density distribution
in such a confined geometry has been demonstrated to be a reliable,
basic tool for investigating
the spectrum of collective excitations and the phase diagram~\cite{ketterle,cornell,grimm}.

From this perspective, one-dimensional (1D) gases show their own peculiarities~\cite{menotti02_bose_hydro1d,petrov04_bec_review}. A paradigmatic example is the exactly solvable
Lieb-Liniger gas ~\cite{Lieb-Liniger}, where at infinite contact interaction
strength $g_{1D} \rightarrow \infty$ the many-body excitation spectrum
becomes identical to that of a free Fermi gas, known as the Tonks-
Girardeau (TG) gas ~\cite{girardeau_bosons1d}. The presence of an external
parabolic potential renders the low-lying part of the excitation
spectrum discrete, where the simplest mode to be excited among the low-lying ones after
small instantaneous changes of the trapping frequency $\omega_z$ is
the so called
breathing (or compressional) mode, {\it i.e.} the uniform radial expansion and contraction of the system.
The breathing-mode frequency $\omega_b$  depends on the interaction strength $g_{1D} $,
the number of particles $N$ in the trap, and the gas temperature $T$. It has been previously shown that the
frequency ratio $\omega_b/\omega_z$ presents two crossovers as a function of increasing interaction:
from the value $2$  down to $\sqrt{3}$ while going from non-interacting to weakly interacting regime,
and then back to $2$ after crossing towards the  strongly interacting limit~~\cite{Haller2009}.
Theoretical descriptions based on local density approximation (LDA)~~\cite{menotti02_bose_hydro1d},
time-dependent Hartree method ~~\cite{time-dep-hf}, and diffusion Quantum Monte Carlo simulations~~\cite{gudyma2015}
have been produced following the system across the different regimes.

Here, we focus on the breathing mode
of a one-dimensional dipolar quantum gas, and investigate the influence of both the dipole orientation and of
the interplay between zero (contact) and finite-range (dipolar) interaction
allowing for independent tuning of these two interactions.
Our analysis is based on a sum-rule approach~~\cite{menotti02_bose_hydro1d} that allows to extract the breathing mode
frequency from ground-state density profiles obtained after solving the stationary generalized Gross-Pitaevskii equation.
The latter is generalized by replacing the Hartree-term with the energy per unit length of the bulk
quasi-one dimensional dipolar system, obtained using either a Bethe Ansatz wave-function in a variational calculation~~\cite{de_palo_variational_2020}, or using a perturbative approach.

The results show that when dipolar interactions are attractive,  the system manifests an incipient instability at low-density and a sharp minimum is found in the breathing mode which is very peculiar of that finite-range interaction. In the repulsive regime an extension of the stability regime is instead observed. A good agreement with the experimental
results reported in Ref.~\onlinecite{Lev2020,Lev_private} is  also found.

The paper is organized as follows. We introduce the model Hamiltonian and the generalized Gross-Pitaevskii equation in Sec.~\ref{sec:I}. Then in Sec.~\ref{sec:II} we discuss the equation of state by separating the short-range terms from the soft dipolar long range interaction in the single-mode approximation. In Sec.~\ref{sec:III} we present the results for the breathing mode by discussing the case of the repulsive and attractive interactions and follow
the evolution of this quantity on varying the dipoles orientation $\theta$.
Finally in Sec.~ \ref{sec:conclusions} we give conclusions and discuss perspectives.

\section{The model and the generalized Gross-Pitaevskii equation}
\label{sec:I}

In highly elongated traps the atomic motion in the plane transverse to the longitudinal direction is described by the Hamiltonian:
\begin{equation}\label{eq:transverse-modes}
H_\perp=\frac{p^2_x+p^2_y}{2 m}+\frac{m \omega^2_\perp}{2} (x^2+y^2),
\end{equation}
where $m$ is the mass particle and $\omega_\perp$ is the confining harmonic oscillator frequency. When the frequency
$\omega_\perp \gg \omega_{ho}$ is sufficiently larger than the longitudinal trapping frequency  $\omega_{ho}$,
the many-body wavefunction of the atoms can be projected on the ground-state manifold of the
Hamiltonian~\eqref{eq:transverse-modes} ~~\cite{dunjko_bosons1d}.  This the so-called single-mode approximation (SMA).

The effective Hamiltonian in the projected subspace depends only on the coordinates along the $z$ axis.
Its expression is~~\cite{Reimann,santos_cond_07}
\begin{eqnarray}
\nonumber
H_{1D}&=&-\frac{\hbar^2}{2 m} \sum_i \frac{\partial^2}{\partial z_j^2}
+g_{1D}\sum_{i<j} \delta(z_i-z_j)\\
&+&\sum_i V_{ext}(z_i) +\sum_{i<j} V_{Q1D}(z_i-z_j),
\label{eq:orig_Ham}
\end{eqnarray}
where $V_{ext}(z)= \frac{1}{2}m \omega^2_{ho} z^2$ is the potential energy of the parabolic trap
along the longitudinal $z$-direction,  $g_{1D}$ is the contact interaction coming from Van der Waals or other
short-ranged interactions, and the effective 1D dipole-dipole interaction $V_{Q1D}(z)$ in the single-mode
approximation reads:~~\cite{Reimann}
\begin{equation}
\label{eq:vq1d}
V_{Q1D}(z/l_\perp)=V(\theta)
\left[ V^{1D}_{DDI} \left(\frac{z}{l_\perp}\right)
-\frac{8}{3}\delta \left(\frac{z}{l_\perp}\right) \right],
\end{equation}
where
\begin{equation}
\label{eq:v_theta}
V(\theta)=\frac{\mu_0\mu^2_D}{4 \pi}\frac{1-3 \cos^2{\theta}}{4 l^3_\perp}
\end{equation}
encodes the sign and the effective strength of the interaction
driven by the vacuum magnetic permeability $\mu_0$, the magnetic dipolar moment $\mu_D$ of the given atomic species, the angle $\theta$ between the dipoles orientation and the longitudinal $z$-axis, and the transverse oscillator length  $l_\perp=\sqrt{\hbar/(m \omega_\perp)}$.
The adimensional form of effective 1D dipolar potential $V^{1D}_{DDI}$ is :

\begin{eqnarray}
\nonumber
&&V^{1D}_{DDI} \left(\frac{z}{l_\perp}\right)=-2\left| \frac{z_i-z_j}{l_\perp}\right|
+\sqrt{2\pi} \left[ 1+ \left(\frac{z_i-z_j}{l_\perp}\right)^2 \right]
\\
&&e^{\left(\frac{z_i-z_j}{l_\perp}\right)^2/2} \erfc \left[\left| \frac{z_i-z_j}{\sqrt{2} l_\perp} \right| \right].
\label{V1d_dd1}
\end{eqnarray}

\noindent In the${}^{162}$Dy case relevant to current experiments~~\cite{Lev2020}, $\mu_D=9.93\mu_B$~~\cite{tang2018}.

At zero temperature, the Gross-Pitaevskii theory~~\cite{pitaevskii1961,gross1963,pitaevskii_becbook} provides a good description of weakly-interacting three dimensional atomic Bose-Einstein condensates, yet it requires modifications either with strong interactions  or  reduced dimensionality.
In the original  form,  without dipolar interaction, the energy functional in the Gross-Pitaevskii approximation
is~\cite{pitaevskii1961,gross1963}
\begin{eqnarray}
\!\!\!F_{GP}\!=\!\!\int\!\!dz\!\left[\frac{\hbar^2}{2m} \nabla \phi\! \nabla \phi^*
\!+\!(V_{ext}(z)\! -\!\mu)|\phi|^2 \!+\!\frac{g_{1D}}{2}|\!\phi|^4 \right]\!\!,
\label{eq:gpe-hartree}
\end{eqnarray}
where $\phi(z,t)$ is the BEC order parameter, $n(z,t)=|\phi(z,t)|^2$ is the boson density, and $\mu$ the chemical potential.
In one dimension and in the case of hard-core bosons~\cite{girardeau_bosons1d},  Kolomeisky \emph{et al.} have proposed a
modification of the Gross-Pitaevskii equation to describe the Tonks-Girardeau regime~\cite{kolomeisky2000}, where  the Hartree term
$g_{1D} |\phi|^4/2$  is replaced by the energy density of the hard core boson (or free spinless fermion~\cite{girardeau_bosons1d,Minguzzi2001}) gas \emph{i.e} ${\hbar^2 \pi^2 |\phi|^6}/({6m})$.  Such approach can be viewed as taking the classical limit in the bosonized Hamiltonian of spinless fermions with quadratic dispersion~\cite{bettelheim_quantum_2008}.
Afterwards, different proposals~\cite{dunjko_bosons1d,ohberg_dynamical_2002,oldziejewski_strongly_2019} have been offered to cover  both the weakly and the strongly interacting regimes, one of them amounts to replace
the Hartree term with an energy-density functional~\cite{dunjko_bosons1d,ohberg_dynamical_2002} for the Lieb-Liniger gas that interpolates between the Hartree and the Tonks-Girardeau limits (see App.~\ref{GGE_LL}).
Indeed, in one dimension, the Lieb-Liniger gas is  integrable by the Bethe Ansatz technique~\cite{lieb_bosons_1D,lieb_excit} and an exact expression of the ground-state energy as a function of the boson density is available.

The  ground-state energy density of the Lieb-Liniger gas reads
\begin{equation}
\label{eq:eneden_ll}
e_{LL}(n)=\frac{\hbar^2}{2 m} n^3 \epsilon_{LL}(n); 
\end{equation}
where $\epsilon_{LL}(n)$ is an adimensional function that can be obtained from the Bethe Ansatz solution~\cite{lang2017,ristivojevic2019,marino2019}.
Using the ground-state energy~(\eqref{eq:eneden_ll}) in the generalized GPE has been shown to reproduce~\cite{ohberg_dynamical_2002} the results of the hydrodynamic approach~\cite{menotti02_bose_hydro1d,petrov04_bec_review} for the lowest breathing mode (see Appendix \ref{GGE_LL} for details).

Along these lines, in this work we replace the Hartree-term in the Gross-Pitaevskii equation~(\eqref{eq:gpe-hartree}) with the energy per unit length of the bulk quasi-one dimensional dipolar system
\begin{equation}
\label{eq:eneden}
e(n)=\frac{\hbar^2}{2 m}  n^3 \epsilon(n), 
\end{equation}
where $\epsilon(n)$ is obtained using either a Bethe Ansatz wave-function in a variational calculation~\cite{de_palo_variational_2020}, or a perturbative approach that we detail in the next section.

The approximation to the energy functional now reads
\begin{equation}
F_{GP}\!=\!\!\int\!\!dz\!\left[\frac{\hbar^2}{2m} \nabla \phi\! \nabla \phi^*
\!+\!(V_{ext}(z)\! -\!\mu)|\phi|^2 \!+ e(|\phi|^2) \right]\!\!,
\label{eq:fggpe}
\end{equation}
yielding the equation of motion~\cite{ohberg_dynamical_2002,oldziejewski_strongly_2019} for $\phi(z,\tau)$, $i \hbar\partial_\tau \phi = {\delta F_{GP}}/{\delta \phi^*}$, \emph{i.e.}
\begin{equation}
\label{eq:ggpe}
i \hbar \partial_\tau \phi=\left[ -\frac{\hbar^2 \nabla^2}{2m} +(V_{ext}(z)-\mu)
+ \frac{1}{\phi}\frac{\delta e(|\phi|^2)}{\delta \phi^*} \right]\phi,
\end{equation}
with the wave function normalized to the number $N$ of atoms in the trap, $N=\int dz |\phi(z)|^2 $.

\section{Equation of state}
\label{sec:II}
We start our analysis by recalling the method used in Ref.~\onlinecite{tang2018} to reduce the system with dipolar interaction~(\ref{eq:vq1d})
to an integrable Lieb-Liniger model. First, in the Hamiltonian (\ref{eq:orig_Ham}) all the short-range contact
interactions are isolated. Then, besides the van der Waals $g_{1D}$ and the contact interaction in Eq.~(\eqref{eq:orig_Ham}),
a contact term $A V(\theta)$ that effectively accounts for  the short-range part of the
interaction $V^{1D}_{DDI}(r)$ is added.
The effective Lieb-Liniger Hamiltonian reads
\begin{eqnarray}
\label{ham:ll_q1d}
&&H^{LL}_{Q1D}=-\frac{\hbar^2}{2m}\sum_i \frac{\partial^2}{\partial x_i^2}\\
&&+\left[ g_{1D}+V(\theta)(A-\frac{8}{3})l_\perp \right]\sum_{i<j} \delta(x_i-x_j),
\nonumber
\end{eqnarray}
where
the normalized strength of the short-range part of the interaction can be  approximately taken as
$
A=\int^{\sqrt{2\pi}}_{\sqrt{2 \pi}} du V^{1D}_{DDI}(u) \simeq 3.6 $, in the single-mode approximation and independently of
the density of atoms~\cite{tang2018}. The  nonzero $A$ takes care
of the shortest-ranged part ($|z|<\sqrt{2\pi} l_\perp$) of the dipolar potential~(\ref{V1d_dd1}) leaving the longer-ranged $~\sim 1/z^3$
integrability-breaking tail as a possible perturbation.

Taking $A=0$ would amount to neglect the short-range part of
the dipolar interaction~(\ref{V1d_dd1}) and thus approximate repulsive or attractive  dipolar interactions with an attractive or repulsive contact interaction, respectively~\cite{Reimann}. Obviously, such an approximation is unphysical. 
The effect of making $A>0$ is to counterbalance the
attractive contact term coming from the single-mode approximation. When $A > 8/3$,
stability is enlarged in the repulsive case, while in the attractive case instability can be obtained for  $g_{1D}$ not sufficiently repulsive.

A reliable estimate of $A$ can be determined via a variational Bethe-Ansatz (VBA) wavefunction
approach~\cite{de_palo_variational_2020}, where, for each density, this effective contact interaction is determined by the minimization
of the energy per particle using the Bethe-Ansatz wavefunction of the Lieb-Liniger model as trial wavefunction.

The dimensionless coupling $\gamma$ for the Lieb-Liniger Hamiltonian defined in (\ref{ham:ll_q1d}) is
\begin{eqnarray}
\label{eq:eff_gamma}
\gamma&=&\frac{1}{n}\frac{m}{\hbar^2}g_{Q1D}(\theta)=
\frac{2}{n a_{Q1D}}\\
\nonumber
&=& \frac{2}{n}\left[ -\frac{1}{a_{1D}}+\frac{a_d}{l^2_{\perp}}
\frac{1-3\cos{\theta}^2}{4} \left(A-\frac{8}{3}\right) \right],
\label{eq:eff_gamma}
\end{eqnarray}
where $g_{1D}=-2 \hbar^2 /(ma_{1D})$ and with $a_d=\mu_0\mu^2_D m/(8\pi \hbar^2)$, the dipolar length.
In this work we will focus on the region where $ a_{1D} < 0$, so that the contact interaction strength
$g_{1D}$ is positive.

In previous modelizations~\cite{tang2018},  the basic assumptions were that (i) $A$ was independent of the density and
 the scattering length $a_{1D}$, and (ii) the tail of the dipolar interaction was negligible.
To start with, let us  include the tail of $V^{1D}_{DDI}(z/l_\perp)$ by means of a perturbative approach.

We write the original Hamiltonian~(\ref{eq:orig_Ham}) as the sum of the
integrable Lieb-Liniger Hamiltonian~(\ref{ham:ll_q1d}) and a  correction term $\delta V$
\begin{eqnarray}
H&=&H^{LL}_{Q1D}(\gamma)+\sum_{i<j}\delta V(z_i-z_j),  \\
\delta V(z)&=&V(\theta) \left[V^{1D}_{DDI}(z_i-z_j)-A l_\perp \delta(z_i-z_j)\right].
\end{eqnarray}
In order to estimate the effect of the interaction $\delta V(z)$,
we resort to perturbation theory (PT). In particular, we will consider two benchmark values for A, $A=3.6$ as in Ref.~\onlinecite{tang2018}, and
$A=0$, which amounts to treat the whole $V(\theta) \left[V^{1D}_{DDI}(z_i-z_j) \right]$ at perturbative
level.
At first order, the energy per $N$ particles is:
\begin{equation}
\frac{E_{pt}}{N}=\epsilon_{pt}(n) \simeq \frac{E_{LL}(\gamma)}{N}+\frac{n}{2} \int dz \delta V(z) g_{LL}(z), 
\label{eq:ham_pt}
\end{equation}
with $E_{LL}$ the Lieb-Liniger ground-state energy for the Hamiltonian (\ref{ham:ll_q1d}) evaluated at $\gamma$,
while $g_{LL}(z)$ is the pair correlation function~\cite{caux_density,cherny2008} of the Lieb-Liniger gas.
Using (\ref{eq:ham_pt}), we obtain an equation of state $e_{PT}(n)$ that depends on the chosen $A$,
besides $a_d/l_\perp,|a_{1D}|/l_\perp$ and $\theta$.

In the rest of the paper, we compare the results obtained  with the following three approximations for the equation of state:
from $\varepsilon_{LL}(n)$, perturbation theory based on \eqref{eq:ham_pt} with $A=0$ and $A=3.6$, and variational
Bethe Ansatz as in Ref.~\onlinecite{de_palo_variational_2020}, that gives a variational estimate for
the ground-state energy independent of the  approximation~\cite{tang2018} chosen for $A$.

\begin{figure}[h]
\begin{center}
\includegraphics[width=90.mm]{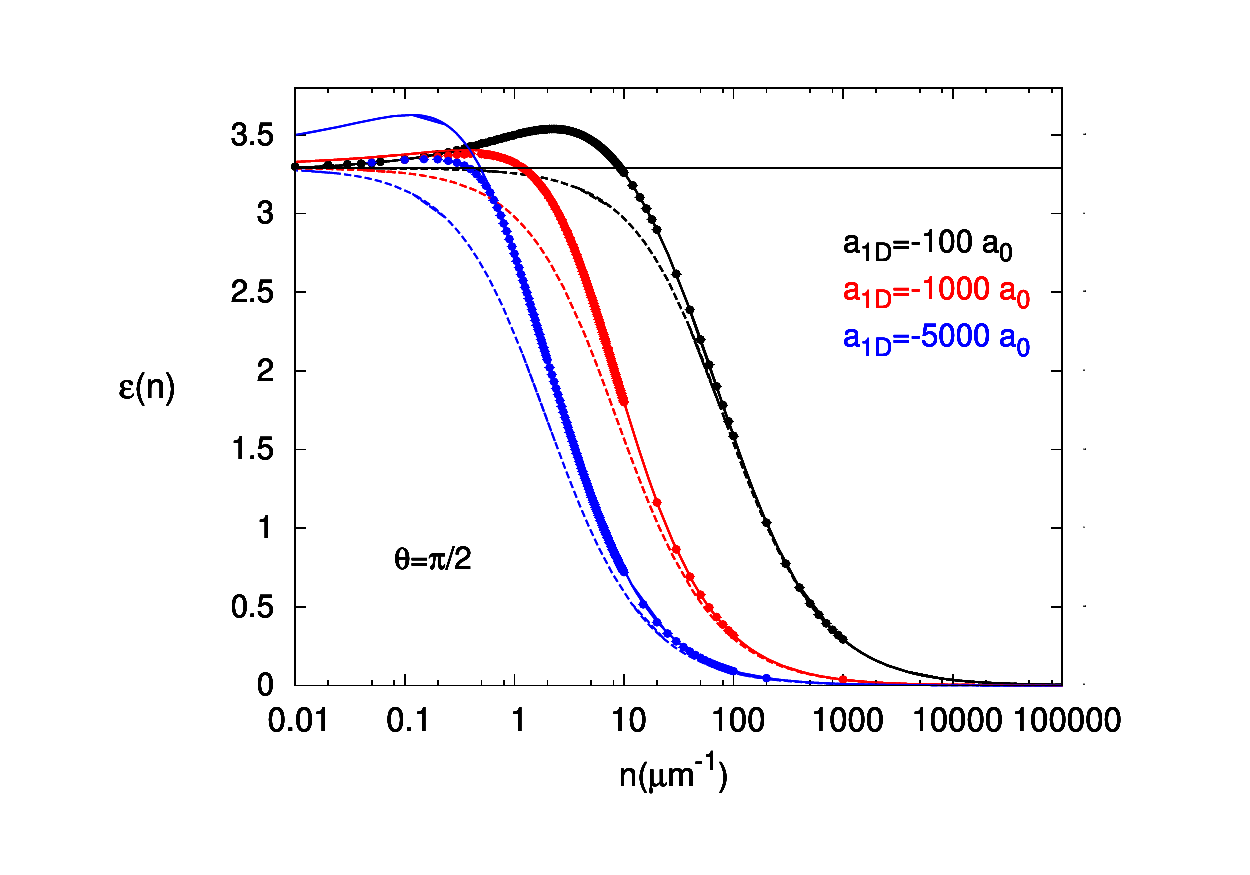}
\end{center}
\caption{Energy per unit length $\epsilon(n)$ in units of ${\hbar^2 n^3}/({2m})$ within three different approximations:
$\epsilon_{LL}(n)$ with $A=3.6$ (dashed lines); perturbation
theory using $A=3.6$ (solid lines) and variational Bethe Ansatz (solid dots). The results are shown
for three selected scattering lengths $a_{1D}=-100 a_0,-1000 a_0 $ and $-5000 a_0$, respectively
the black, red and blue data and for $\theta=\pi/2$.}
\label{fig:cfr_ene}
\end{figure}

In Fig.~\ref{fig:cfr_ene} we show the energies $\epsilon(n)$ within three different approximations:
$\epsilon_{LL}(n)$ using Eq.~\eqref{ham:ll_q1d} and $A=3.6$,  $\epsilon_{PT}(n)$ with $A=3.6$,
and finally the variational Bethe Ansatz. Results are shown for three selected scattering lengths
$a_{1D}=-100 a_0,-1000 a_0 $
and $-5000 a_0$ and for $\theta=\pi/2$, i.e. for repulsive interaction.
We choose $a_d=195 a_0$,  $l_\perp=57.3  \mathrm{nm}$ and $a_{ho}=24000 a_0$, to make contact with recent
experimental works~\cite{tang2018,Lev2020}.
For small scattering lengths, the equations of state from the perturbative approach using $A=3.6$ and from the
variational Bethe Ansatz are in good agreement with each other.
Equations of state  within these two approximations, visibly depart from $\epsilon_{LL}$ based on Eq.~\eqref{ham:ll_q1d}
at small and intermediate densities and on increasing the scattering length as expected since the dipolar
interactions becomes more dominant.

\begin{figure}[h]
\begin{center}
\includegraphics[width=90.mm]{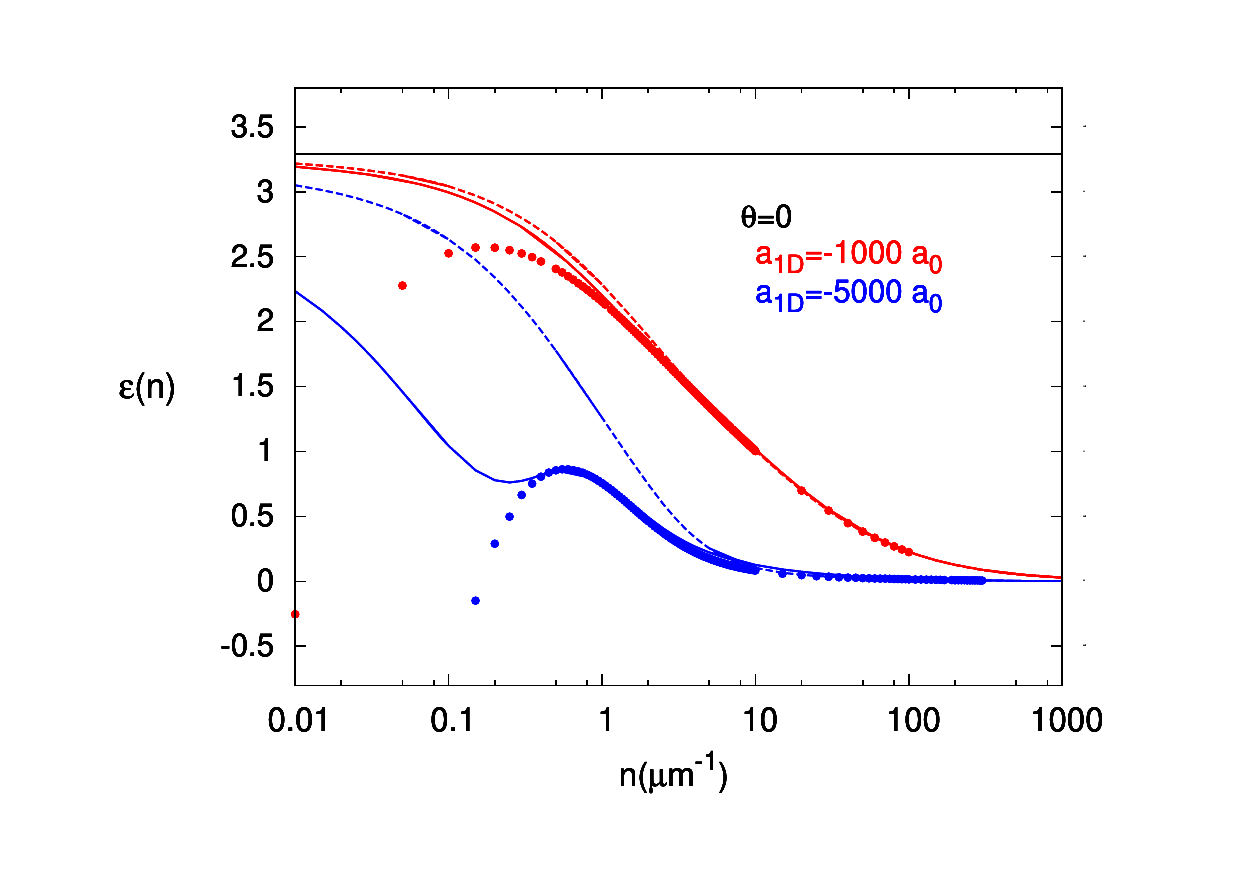}
\end{center}
\caption{Energy per unit length $\epsilon(n)$ in units of ${\hbar^2 n^3}/({2m})$ within three different approximations:
using perturbation theory with $A=3.6$ (solid lines)
and using $A=0.0$ (dashed lines) compared with those within the variational
Bethe Ansatz (solid dots). Data are shown for two selected scattering lengths $a_{1D}=-1000 a_0 $ and $-5000 a_0$, represented by
the red and blue data points, respectively. Results are for $\theta=0$.}
\label{fig:cfr_ene_zero}
\end{figure}

The typical situation for the attractive interaction, ie. for $\theta=0$, is displayed in Fig.~\ref{fig:cfr_ene_zero}, where we show
the data for $a_{1D}=-1000 a_0 $ and $-5000 a_0$, and compare the energy results coming from variational Bethe Ansatz, perturbation theory using $A=3.6$ for all densities, and using
$A=0.0$, that is treating the whole dipolar interaction as a perturbation.
At low and intermediate densities, energies obtained within PT with $A=0$
largely deviates from VBA results and within themselves.
Only at very large densities these differences reduce, and
results from PT with $A=0$ are closer to the variational
results.
It should be kept in mind that those differences are strongly reduced by the $n^3$ factor in the energy per volume $E/V=n^3 \epsilon(n)$.

These results emphasize that using a single effective contact interaction $A$, independent of  both density and scattering length,
can yield  an inaccurate equation of state, especially with attractive dipolar interactions. The VBA approach, by optimizing the parameter $A$ to minimize the ground state energy, 
takes care of these uncertainties.
The relevance of such differences for the calculation of breathing mode frequency in a trapped system will be considered in the next section.

\section{The breathing mode}
\label{sec:III}

We evaluate the frequency of the lowest radial compressional oscillation by a sum-rule approach
~\cite{menotti02_bose_hydro1d}, that allows to compute the breathing mode frequency
from ground-state density profiles obtained by solving
the stationary generalized Gross-Pitaevskii equation using standard imaginary time evolution algorithms~\cite{kumar2015}.
The breathing mode $\omega_b$ is obtained as the response of the gas to a change of the trap
frequency $\omega_{ho}$:

\begin{equation}
\omega^2_b= -2 \langle \sum^N_{i=1} z^2_i\rangle \left[\frac{\partial \langle \sum^N_{i=1} z^2_i\rangle}
{\partial \omega^2_{ho} }\right]^{-1}. 
\end{equation}
It is convenient (see App.~\ref{app:inhom-tll}) to study the evolution of the breathing mode as a function of
$\Lambda=N a_{1D}^2/a_{ho}^2$, with $N$ the number of particles in the trap.
By solving the time-dependent generalized Gross-Pitaevskii equation, we have verified that, after initially exciting the mode by external radial compression of the trap, in the limit $|a_{1D}|\rightarrow 0$
$(\omega_b/\omega_{ho})^2=4$.

\begin{figure}[h]
\begin{center}
\includegraphics[width=90.mm]{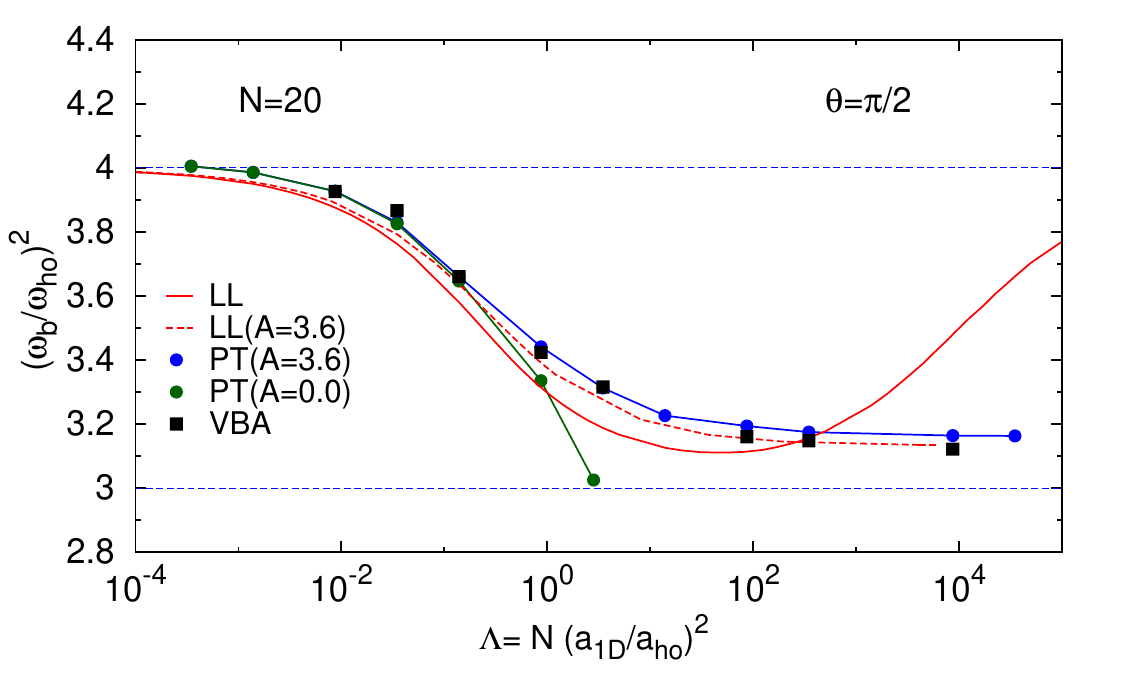}
\end{center}
\caption{
Squared breathing mode frequency over trapping frequency squared $\omega^2_b/\omega^2_{ho}$ as a function of
$\Lambda=N (a_{1D}/ a_{ho})^2$, using different approximations. All results are for $N=20$ and lines are only a guide to the eyes.
Red, dark-green and blue solid curves are estimates using the Lieb-Liniger model, perturbation theory with
$A=0$ and with $A=3.6$, respectively. The dashed red curve represents $\omega^2_b/\omega^2_{ho}$ calculated using
Eq~\eqref{ham:ll_q1d} with $A=3.6$.
the black solid squares are estimates based on the Variational Bethe-Ansatz equation of state. }
\label{fig:exp_pi2}
\end{figure}

We estimate the breathing mode using the different approximations described above, starting from the
case $\theta=\pi/2$ (see Fig.~\ref{fig:exp_pi2}).

In the region of small scattering lengths $|a_{1D}|$, the breathing modes are dominated
 by the van der Waals repulsive contact interaction and dipolar interactions are marginally relevant:
all the approximations, even completely neglecting the dipolar interaction, predict similar results.
On increasing $|a_{1D}|$, apart from using PT with $A=0$ that fails when $g_{q1d}(\pi/2)$ becomes negative,
all the other approximations shown in Fig.~\ref{fig:exp_pi2} are very close to each other.
The important effect of dipolar interaction becomes visible for very large $|a_{1D}|$ values, where it
enlarges the region of stability and, for $|a_{1D}|\rightarrow \infty$, the breathing-mode frequency
reaches a plateau.
This behavior can be already obtained within the approximation of ~\cite{tang2018}  
since with $A=3.6$, according to Eq.~\eqref{ham:ll_q1d},  $g_{q1d}(\pi/2)$ saturates in that limit.

We note that when $\Lambda < 10^3$, the estimates of the breathing mode frequency from both
the PT using $A=3.6$ and the VBA are compatible with the one obtained by dropping entirely the dipolar interaction. 
This last modelization however would predict that the breathing mode reaches the non-interacting
limit at large $\Lambda>10^4$, i.e. $(\omega_b/\omega_{ho})^2 \rightarrow 4$, at variance with the other
two approximations that predict a plateau at a lower  $(\omega_b/\omega_{ho})^2 \simeq 3.2$, hinting at the persistence of interactions.

\begin{figure}[h]
\begin{center}
\includegraphics[width=90.mm]{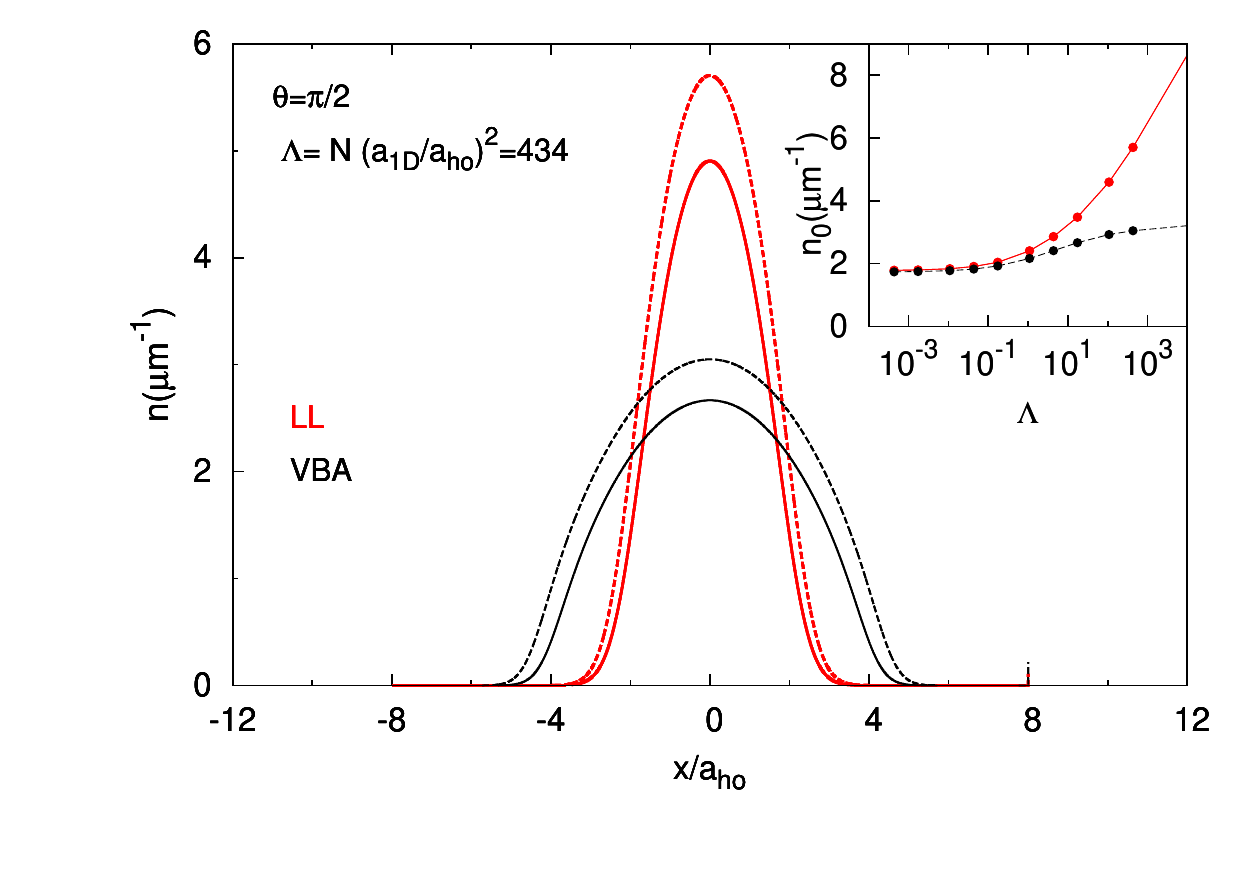}
\end{center}
\caption{Density profile for a system of $N$ dipoles in a trap using two different equations
of states at $\Lambda=434$, namely the Lieb-Liniger (LL) (red lines) and VBA (black lines).
Dashed and solid lines refer to the case with $N=25$ and $N=20$ particles in the trap, respectively.
Results are for $\theta=\pi/2$.
In the inset, we show the density at the center of the trap as a function of $\Lambda$ for $N=25$,
the red and black solid dots representing estimates using the LL and VBA equations of state, respectively.
Lines joining dots are only guide to the eye.}
\label{fig:den_pi2}
\end{figure}

At large $|a_{1D}|$, even when the
predictions for the breathing mode are all compatible, we can trace a difference in the density profile, as illustrated in
Fig.~\ref{fig:den_pi2} at $\Lambda=434$, where we show it using VBA and LL. 

Using either $\epsilon_{VBA}(n)$ or $\epsilon_{PT}(A=3.6,n)$ yields a density at the center of the trap
that ranges from
$ \simeq 2.6$ to  $3.1 \mu \mathrm{m}^{-1}$ (see the black curves in the main panel of Fig.~\ref{fig:den_pi2}),
while the density value at the center of the trap is almost doubled for the Lieb-Liniger gas without dipolar
interaction described by $\epsilon_{LL}(n)$.
The first estimates are in agreement with the averaged density at the center of trap as measured in
Ref.~\onlinecite{tang2018}.
In the inset we show the variation of the density at the center of the trap as a function of $\Lambda$
for the Lieb-Liniger equation of state and the VBA. The values become notably different for $\Lambda \agt 10$, whereas (see Fig.~\ref{fig:exp_pi2}) the behavior of the breathing mode becomes qualitatively different for the two approximations only for $\Lambda \agt 500$.  The behavior of the density profiles shows that  the physics of the system at large scattering lengths is different in the presence of repulsive dipolar interactions, and could be used as
a sensitive indicator together with the frequency of the breathing mode.

Turning to  the attractive case, i.e. $\theta=0$, on increasing $|a_{1D}|$ the $g_{Q1D}(\theta)$
in Eq.~\eqref{eq:eff_gamma} becomes rapidly small and negative, and the key issue is to what extent the
system of dipolar gas is still stable against possible collapse ~\cite{McGuire64}, the formation of a solitonic/droplets
phase~\cite{oldziejewski_strongly_2019} or a gas/droplet coexistence ~\cite{Kora_2020}.
The  predictions for the breathing modes are qualitatively different from the repulsive case, since
both the VBA and the estimates with $A=3.6$, with or without correction to the first order,
predict that for $\Lambda > 2$ the breathing mode rapidly decreases to reach a
minimum with $\omega^2_b/\omega^2_0 < 3$, after which it rapidly increases until the overall effective interaction
becomes negative. Due to negative $g_{q1d}$, also the LL model using $A=3.6$ (Eq.~\eqref{ham:ll_q1d}) predicts an instability, yet at a higher value of
$\Lambda$ than the two other approximations with $ 3<\omega^2_b/\omega^2_{ho}<4$.

\begin{figure}[h]
\begin{center}
\includegraphics[width=90.mm]{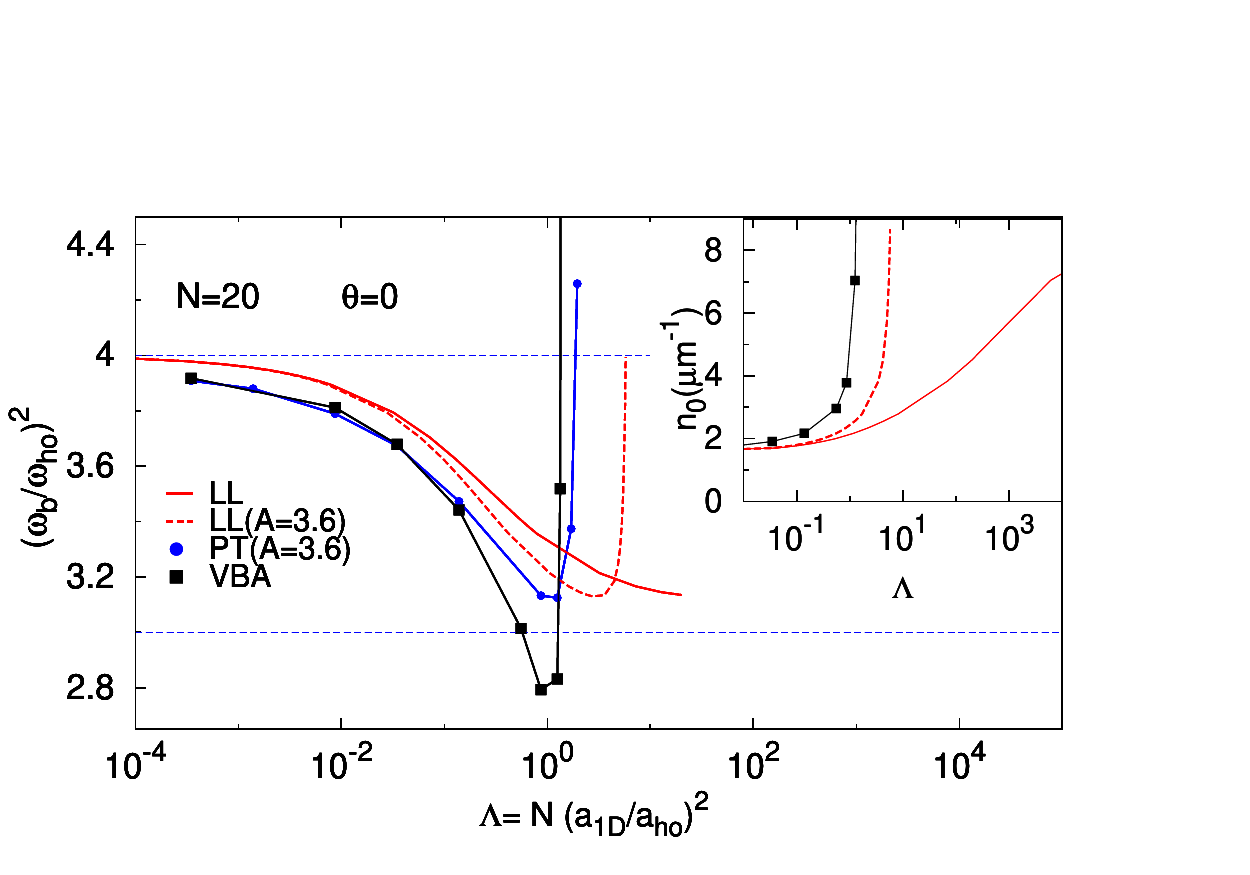}
\end{center}
\caption{Squared breathing mode frequency over trapping frequency squared $\omega^2_b/\omega^2_{ho}$ as a function of
$\Lambda=N (a_{1D}/ a_{ho})^2$, using different approximations.
The red solid and dashed lines represent the breathing modes
estimates after using the Lieb-Liniger (LL) model without dipolar interaction and using $A=3.6$ in Eq.~\eqref{ham:ll_q1d}, respectively.
Blue solid curves refer to estimates based on the perturbation theory with
$A=3.6$ while
black solid squares are based on the Variational Bethe-Ansatz (VBA). 
All results are for $N=20$ and lines are only guides to the eye. In the inset, the density at the center of the trap is shown as a function of $\Lambda$, with
the same legend as in the main panel. }
\label{fig:exp_zero}
\end{figure}

The important discrepancy for $\Lambda>1$ between the $A=3.6$ approximation and its first-order correction, is suggesting a
breakdown of this approximation. If we contrast with the VBA, we observe that the latter predicts a deeper minimum of the breathing
mode frequency than the $A=3.6$ approximation, even with first-order corrections.
Comparing the densities at the center of the trap (see inset of Fig.~\ref{fig:exp_zero}), we note that differences in density
are becoming noticeable already for $\Lambda \sim 0.1$, suggesting again that the density profile is more sensitive to the
presence of dipolar interaction than the breathing mode. In any case, all the approximations confirm that we are approaching
an instability at intermediate values of $\Lambda\sim 1$. Of course, only a direct comparison with experimental data could permit to identify which approximation is the most suitable for other predictions, as we will see later on.

Comparing the repulsive and attractive cases, we see that attractive dipolar interactions produce stronger qualitative
effects on the behavior of the breathing mode or on the density profile at a given $a_{1D}$.
In addition, differences between the VBA and the perturbation theory with $A=3.6$ are also more significant in the presence of repulsive interactions.
Such observation is in agreement with the behavior of the energy density represented in Figs.~\ref{fig:cfr_ene} and \ref{fig:cfr_ene_zero}, where the differences between the approximations manifest themselves for lower $|a_{1D}|$ in the attractive case.

\begin{figure}[h]
\begin{center}
\includegraphics[width=90.mm]{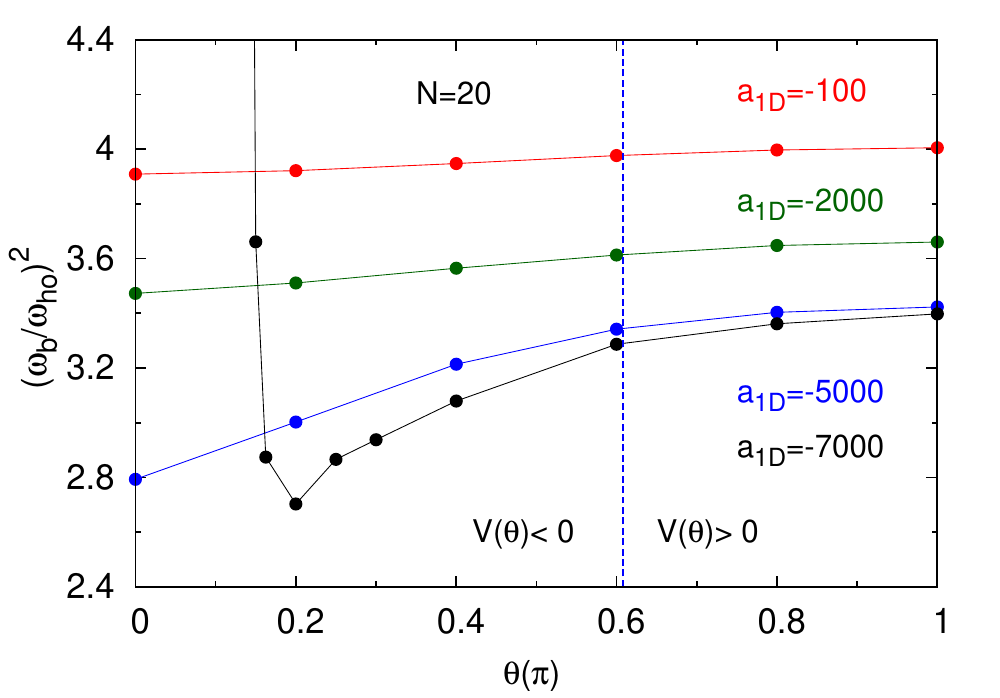}
\end{center}
\caption{Squared breathing mode frequency over trapping frequency squared $\omega^2_b/\omega^2_{ho}$ as a function of
the polarization angle $\theta$ for selected values of the scattering length $a_{1D}/a_0=-100, -2000,-5000,-7000$, represented by
red, dark-green, blue and black solid dots, respectively. The solid lines joining the data
are only guides to the eye. The vertical blue dashed line splits the regions with negative $V(\theta)< 0$ (left) and positive $V(\theta)>0$ (right).
Data refer to estimates based on the Variational Bethe-Ansatz equation
of state.}
\label{fig:bm_theta}
\end{figure}

Finally we show in Fig.~\ref{fig:bm_theta} the effect of changing the polarization angle $\theta$, while keeping fixed
the scattering length $|a_{1D}|$. For a large range of scattering lengths
the effect of varying the angle is very small and  visible only just before the system becomes unstable. For the largest scattering length, and attractive interaction, the breathing mode rapidly grows, signaling the  instability, as previously found.

\begin{figure}[h]
\begin{center}
\includegraphics[width=90.mm]{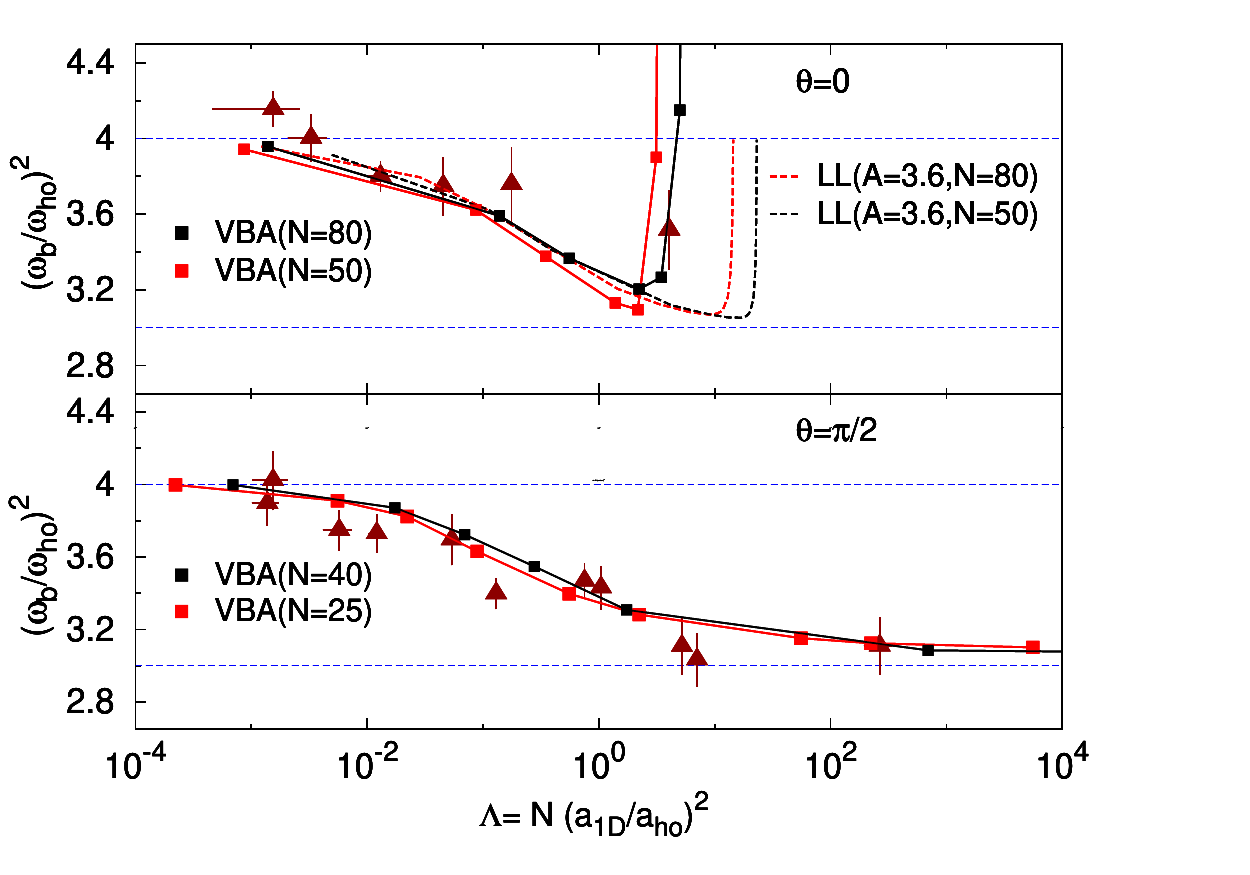}
\end{center}
\caption{
Squared breathing mode frequency over trapping frequency squared $\omega^2_b/\omega^2_{ho}$ as a function of
$\Lambda=N (a_{1D}/ a_{ho})^2$, using different approximations, for the attractive case (upper panel) and the repulsive one (lower panel).
Dark-red triangles represent the experimental data taken from Ref.~\onlinecite{Lev2020,Lev_private}. All results are shown for two values of $N$, $N=50,80$ (upper panel) and $N=25,40$ (lower panel) and lines are only a guide to the eyes.
The dashed black and red lines represent estimates using the Lieb-Liniger model. Black and red solid squares are estimates based on the Variational Bethe-Ansatz (VBA) equation of state.}
\label{fig:exp_all}
\end{figure}

We conclude our discussion by contrasting the proposed approximation with the experimental data from Ref.~\onlinecite{Lev2020,Lev_private},
as shown in Fig.~\ref{fig:exp_all}.
We note that for the repulsive case ($\theta=\pi/2$) all the experimental points are in very good agreement with the VBA
prediction both with $N=25$ and $N=40$, that are the minimum and maximum number of particles in the trap characterizing the experiment.
Our findings suggest that for $\Lambda\geq 1$ the dipolar interaction is efficient in enhancing the region of stability of the
interacting regime and in inhibiting the increase of the breathing mode frequency towards the non-interacting limit.
The agreement with the VBA predictions is also confirmed for the attractive case ($\theta=0$) where the curve with $N=80$ and $N=50$
agrees well with the experimental points. The comparison with the LL theory, using $A=3.6$ and neglecting the tail interaction
(Eq.~\eqref{ham:ll_q1d}) clearly shows that despite its correct qualitative behavior the VBA description is needed to make contact
with experimental findings.
The comparison with the experiments clearly indicate that the system crosses over an instability point for $\Lambda$ of the order 1.
Whether this instability is due to the formation of simple bound states ~\cite{McGuire64} or droplets formation
~\cite{oldziejewski_strongly_2019} needs further investigations, in particular from the experimental point of view.

\section{Conclusions}
\label{sec:conclusions}

In conclusion, we have considered the energy density of a gas of dipolar bosons in a tight transverse trapping using either the approximation of Ref.\onlinecite{tang2018}, supplemented either by first order perturbation theory, or a variational approximation~\cite{de_palo_variational_2020}. We have found that in the case of repulsive dipolar interactions, the two approaches were in good agreement with each other. We have used the energy densities under the different approximations to predict the breathing mode frequencies of the trapped dipolar gas.  When dipolar interactions become attractive, the results of the two approximations become quite different, especially at low density. This gives rise to notable differences in the frequency of the breathing mode, the variational method giving a stronger dip before the instability. In all cases, observing the effect of the dipolar interaction requires to weaken enough the contact interaction that is competing with it. Contrasting with experimental results, we have shown that the variational predictions are especially compatible with the present measurements~\cite{Lev2020, Lev_private}. However, except than in the attractive case, the experimental results are also compatible with a pure contact interaction.

As already noticed in our previous work~~\cite{citro06_dipolar1d}, this can be considered as a further proof that to all relevant purposes the nature of $1/r^3$ power-law interactions in 1D, in the ground state repulsive branch, can be viewed as short-range interactions~\cite{dalmonte2010}. It would be worthwhile that future experiments attempt to explore the region with $\Lambda>400$ in the repulsive regime, where deviations for the pure contact interaction are expected, and the range $0.1<\Lambda<1$ in the attractive case, where deviations from the pure contact interaction are maximal, and the difference between the two approximations considered here are the most visible.

\begin{acknowledgments}
We thank Benjamin Lev and his group for enlightening discussions and for private communication of data and Chiara Menotti for useful discussions.
\end{acknowledgments}

\appendix
\section{Generalized Gross-Pitaevskii for the Lieb-Liniger trapped system }\label{GGE_LL}

\begin{figure}[h]
\begin{center}
\includegraphics[width=90.mm]{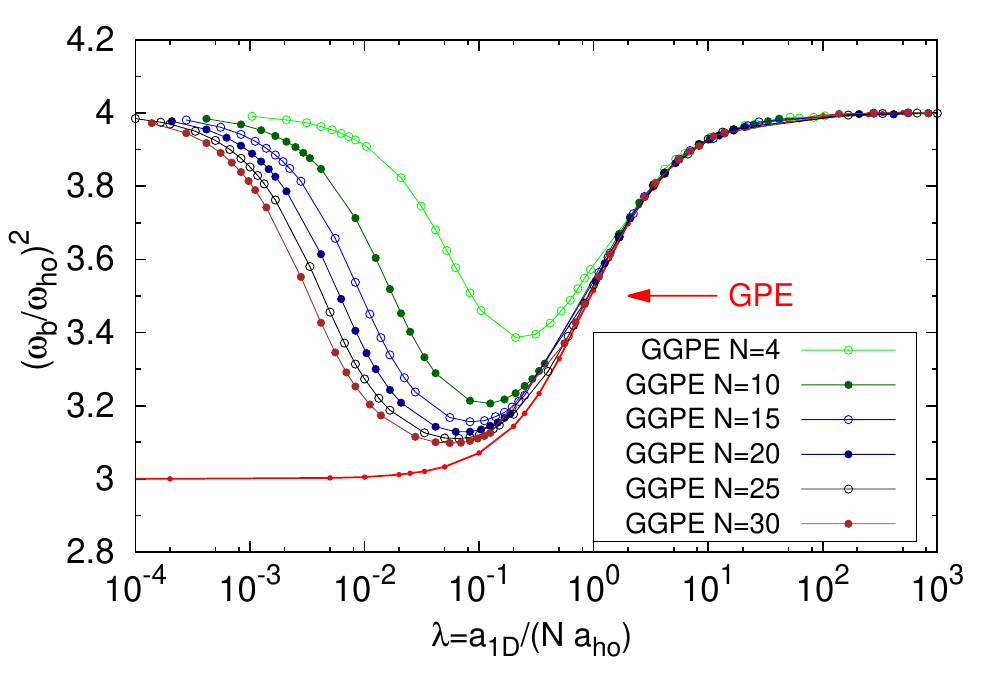}
\end{center}
\caption{Ratio of breathing mode frequency to trap frequency squared $\omega^2_B/\omega^2_{ho}$ as a function of the Hartree parameter $\lambda$ for different numbers
of particles in the trap, namely $N=4,10,15,20,25$ and $30$ using the generalized Gross-Pitaevskii equation
Eq.~\eqref{eq:GGPE}. The red solid line represents the Hartree approximation, independent of the number of particles.}
\label{fig:bm_ll_N}
\end{figure}

We replace the Hartree term in Eq.(\eqref{eq:gpe-hartree}) with the ground-state density energy of the Lieb-Liniger gas
Eq.~(\eqref{eq:eneden_ll}), so that the energy functional and the equation of motion reads:
\begin{eqnarray}
&&F_{GP}=\int dz \left[\frac{\hbar^2}{2m}| \nabla \phi |^2 +
V_{ext}(z) |\phi|^2 +  e(|\phi|^2) \right] \\
\label{eq:GGPE}
&& i \hbar \partial_\tau \phi=\left[ -\frac{\hbar^2 \nabla^2}{2m} +V_{ext}(z) + \frac{1}{\phi}\frac{\delta e(|\phi|^2)}{\delta \phi^*}\right]\phi,
\end{eqnarray}
where
\begin{eqnarray}
&&\!\!\frac{1}{\phi}\frac{\delta e(n)}{\delta \phi^*}=\!\!\frac{\hbar^2}{2m} \left[\!3 n^2 \epsilon_{LL}(\gamma[n])
\! -\!\!\frac{2 n}{a_{1D}}\frac{d \epsilon_{LL}(\gamma[n])}{d \gamma}\right],
\label{A3}
\end{eqnarray}
with $\epsilon_{LL}(\gamma[n])$ the adimensional ground-state density energy functional
for the Lieb-Liniger as for example from the works \onlinecite{lang2017,ristivojevic2019,marino2019}.
We use a normalized wave-function as well as
harmonic-oscillator units, that means that lengths and energies are respectively expressed in units of
$a_{ho}$ and $\hbar \omega_{ho}$, so that:
\begin{eqnarray}
n&=& \frac{N}{a_{ho}}|\psi|^2 \\
\gamma&=&\frac{2 a_{ho}}{N |\phi|^2 a_{1D}}=\frac{2}{N^2 \lambda |\psi|^2} \\
\lambda&=&\frac{a_{1D}}{N a_{ho}},
\end{eqnarray}
where $\lambda$ is the Hartree parameter.
In these units, Eq.~\ref{A3} becomes
\begin{equation}
\frac{1}{\phi}\frac{\delta e(n)}{\delta \phi^*} = \frac{3}{2} N^2 |\psi|^4 \epsilon(\gamma[n])-\frac{1}{\lambda} |\psi|^2 \epsilon'(\gamma[n]),
\end{equation}
and it covers the strong and the weak interaction cases.
Indeed, in the weak-interacting limit
\begin{eqnarray*}
&& \lim_{\gamma \rightarrow 0}
\frac{1}{\phi}\frac{\delta e(n)}{\delta \phi^*}
 \rightarrow  \frac{3}{\lambda} |\psi|^2 -\frac{1}{\lambda} |\psi|^2 = \frac{2}{\lambda} |\psi|^2
\end{eqnarray*}
we recover the usual Gross-Pitaevskii equation
\begin{eqnarray}
i \hbar \partial_t \psi=\left[ \frac{1}{2}\left(-\nabla^2+x^2 \right)+\frac{2}{\lambda} |\psi|^2 +\right] \psi,
\end{eqnarray}
while in the strong interacting limit
\begin{equation*}
\lim_{\gamma \rightarrow \infty} \frac{1}{\phi}\frac{\delta e(n)}{\delta \phi^*} \rightarrow \frac{3}{2} N^2 |\psi|^4 \epsilon(\gamma[n]) = \frac{\pi^2}{2} N^2 |\psi|^4
\end{equation*}
we get back to the proposal from Kolomeisky \emph{et al.} ~\cite{kolomeisky2000}
to describe the Tonks-Girardeau gas
\begin{eqnarray}
i \hbar \partial_t \psi=\left[\frac{1}{2}\left(-\nabla^2+x^2 \right)\!+\!\frac{\pi^2}{2} N^2 |\psi|^2 |\psi|^2 +\right]\psi.
\end{eqnarray}
Results for the breathing modes using this approach, for different number of particles in the trap, are shown in
Fig.~\ref{fig:bm_ll_N} and compared with the usual Gross-Pitaevskii equation (GPE)

\section{Breathing mode of an inhomogeneous Tomonaga-Luttinger liquid}\label{app:inhom-tll}
Here, we briefly recall the relevant parameters to study the evolution of the breathing mode in a trapped Tomonaga-Luttinger liquid~\cite{menotti_stringari,petrov04_bec_review,citro07_luttinger_hydrodynamics}.
The Hamiltonian of the inhomogeneous Tomonaga-Luttinger liquid reads
\begin{equation}
  \label{eq:tll-inhom}
  H=\int_{-R}^{R} \frac{dx}{2\pi} \hbar \left[ u(x) K(x) (\pi \Pi)^2 + \frac{u(x)}{K(x)} (\partial_x \phi)^2\right],
\end{equation}
where~\cite{citro07_luttinger_hydrodynamics}
\begin{eqnarray}
  \label{eq:tll-params}
  u(x) K(x) &=&\frac{\hbar \pi \rho(x)}{m},  \\
  \frac{u(x)}{K(x)}&=&\frac 1 {\pi \hbar} \left(\frac{\partial \mu}{\partial \rho}\right)_{\rho=\rho(x)},
\end{eqnarray}
with $\rho(x)$ the density of atoms at position $x$, $\mu$ the chemical potential, $m$ the mass of a single atom, and $2R$ the dimension of the trapped atomic cloud.
Using the equations of motion method, one obtains~\cite{menotti_stringari,petrov04_bec_review,citro07_luttinger_hydrodynamics}
\begin{eqnarray}
  \label{eq:tll-modes}
  \partial_t^2 \phi &=& u(x) K(x) \partial_x \left(\frac{u(x)}{K(x)} \partial_x \phi\right)\\
  &=& \frac{\rho(x)}{m}  \partial_x \left(\frac{\partial \mu}{\partial \rho } \partial_x \phi\right).
\end{eqnarray}
The breathing modes are obtained by looking for solutions of (\ref{eq:tll-modes}) of the form $\phi(x,t) = \phi_n(x) e^{i\omega_n t}$ subject to the boundary conditions $\phi_n(\pm R)=0$. The local chemical potential in a harmonic trap is fixed by the equation
\begin{equation}
  \label{eq:mu-x-relation}
  \mu(\rho(x)) = \frac 1 2 m \omega_0^2 (R^2-x^2).
\end{equation}
In the case of the Lieb-Liniger gas, the energy per unit length is given by~(\ref{eq:eneden})
\begin{equation}
  \label{eq:lieb-liniger}
  e(\rho)=\frac{\hbar^2 \rho^3}{m} \bar{\epsilon}(\rho a_{1D}),
\end{equation}
therefore it is convenient to use a reduced density $\nu(x)=a_{1D} \rho(x)$, and write $\mu(\rho)=\partial_\rho e(\rho)$ in the form
\begin{equation}
  \mu =\frac{\hbar^2}{m a_{1D}^2} \psi(\nu),
\end{equation}
so that after inverting (\ref{eq:mu-x-relation}) we find
\begin{equation}\label{eq:nu-x}
  \nu = \psi^{-1} \left(\frac{a_{1D}^2 (R^2-x^2)}{a_{ho}^4}\right),
\end{equation}
where we have introduced the trapping length $a_{ho}=\sqrt{{\hbar}/({m \omega_0})}$. If we consider the total number of particles $N$, we have
\begin{eqnarray}
  N&=&\int_{-R}^R \rho(x) dx,
\end{eqnarray}
and injecting (\ref{eq:nu-x}), we find that
\begin{equation}
  \frac{N a_{1D}}{R} = \int_{-1}^1 du  \psi^{-1} \left(\frac{a_{1D}^2 R^2 (1-u^2) }{a_{ho}^4}\right).
\end{equation}
Solving that equation yields
\begin{equation}
  R=\frac{a_{ho}^2}{a_{1D}} G(\Lambda),
\end{equation}
with $\Lambda=N a_{1D}^2/a_{ho}^2$. Introducing the dimensionless variable
$\xi={a_{1D} x}/{a_{ho}^2}$, we can rewrite the density and the chemical potential in the form:
\begin{eqnarray}
  \rho(x)=a_{1D}^{-1} F_1[G(\Lambda)^2 - \xi^2]\\
  \partial_\rho\mu(\rho(x)) = \frac{\hbar^2}{m a_{1D} } F_2[G(\Lambda)^2 - \xi^2],
\end{eqnarray}
and obtain the dimensionless eigenvalue equation
\begin{equation}
  \label{eq:eigen-dimensionless}
F_1[G(\Lambda)^2 - \xi^2] \partial_\xi \left\{F_2[G(\Lambda)^2 - \xi^2] \partial_\xi \phi_n \right\} = -(\omega_n/\omega_0)^2 \phi_n,
\end{equation}
with boundary conditions $\phi_n(\xi=\pi G(\Lambda))=0$. So, in the case of the Lieb-Liniger gas the eigenvalues $(\omega_n/\omega_0)^2$ depend only on the parameter $\Lambda$. Obviously, this is not going to be the case in the dipolar gas where the ground state energy per unit length depends also on the dimensionless ratios $a_{1D}/a_d$, $a_{1D}/l_\perp$ and the angle $\theta$. However, in a limit where the dipolar interaction can be replaced by an effective contact interaction, the same kind of scaling will hold.
%

\end{document}